**Real-time dynamics and cross-correlation gating spectroscopy of free-carrier Drude slow-light solitons**

H. Zhou[1,4,*,†], S.-W. Huang[2,4,*,‡], X. Li[3,4], J. F. McMillan[4], C. Zhang[5], K. K. Y. Wong[5], M. Yu[6], G.-Q. Lo[6], D.-L. Kwong[6], K. Qiu[1], and C. W. Wong[2,4,§]

[1] Key Lab of Optical Fiber Sensing and Communication Networks, University of Electronic Science and Technology of China, Chengdu 611731, China.
[2] Mesoscopic Optics and Quantum Electronics Laboratory, University of California, Los Angeles, CA 90095, USA.
[3] College of Science, National University of Defense Technology, Changsha, Hunan 410073, China.
[4] Optical Nanostructures Laboratory, Columbia University, New York, NY 10027, USA.
[5] Department of Electrical and Electronic Engineering, The University of Hong Kong, Pokfulam Road, Hong Kong.
[6] Institute of Microelectronics, Agency for Science, Technology and Research, Science Park II, 117685, Singapore.

*equal contribution. E-mail: † zhouheng@uestc.edu.cn; ‡ swhuang@seas.ucla.edu;
§ cheewei.wong@ucla.edu.

**Optical solitons – stable waves balancing delicately between nonlinearities and dispersive effects – have advanced the field of ultrafast optics and dynamics, with contributions spanning supercontinuum generation and soliton fission, to optical event horizon, Hawking radiation, and optical rogue waves, amongst others. Here we investigate picojoule soliton dynamics in silicon slow-light photonic-bandgap waveguides under the influence of Drude-modeled free-carrier induced nonlinear effects. Using real-time and single shot amplified dispersive Fourier transform spectroscopy simultaneously with high-fidelity cross-correlation frequency-resolved optical gating at femtojoule sensitivity and femtosecond resolution, we examine the soliton stability limits, the soliton dynamics including free-carrier quartic slow-light scaling and acceleration, and the Drude electron-hole-plasma induced perturbations on Cherenkov radiation and modulation instability. Our real-time single shot and time-averaged cross-correlation measurements are matched with**

**our detailed theoretical modeling, examining the reduced group velocity free-carrier kinetics on solitons at picojoule.**

Introduction

Solitons are stable wave packets that emerge from autonomous balance between dispersive and nonlinear phase shift [1,2], examined in diverse media including various optical fibers [3-8], photonic crystal waveguides [9-11], nonlinear photonic lattices [12], and resonant cavities [13,14]. The formation and propagation of solitons reflect precisely the physical characteristics of the operating regimes, with key dynamics including Raman-induced soliton self-frequency shift [3,4], soliton fission and compression [5,9,10], four-wave mixing [15], optical Cherenkov radiation [16-18], supercontinuum generation [4,6,16], optical rogue waves and fluctuations [19-20], astronomical-distance interaction of solitons [21], self-similarity [22] and event horizon analogs [8], amongst others [23]. At the same time, advances in the silicon photonic platforms, especially with slow-light photonic crystal (PhC) structures, provide a broad parameter space for soliton dynamics with its tailorable dispersion and enhanced Kerr nonlinearity [9-11,24-28]. Intrinsic two-photon absorption (TPA) of single-crystal silicon, when pumped with high peak power near-infrared lasers, drives high density of free carriers and its corresponding Drude plasma dynamics. This has been shown to induce soliton acceleration [10,29-30], soliton compression [31], and soliton fission [32] in photonic crystal waveguides.

Here we advance the dynamical studies of slow-light solitons on-chip, with the inherent Drude electron-hole plasma dynamics, through the simultaneous real-time single shot amplified dispersive Fourier transform (ADFT) and cross-correlation frequency resolved optical gating (XFROG) spectroscopies. First, a quartic slow-light scaling (at 7% the vacuum speed of light) of the free-carrier kinetics is described based on high-fidelity XFROG characterizations of solitary pulses. We provide closed-form analytical solutions, together with the numerical modeling and measurements to describe the self-induced temporal acceleration and self-frequency blue-shift. Second, we extend the description of the Drude electron-hole plasma perturbations onto femtosecond pulse spectrum broadening and optical Cherenkov radiation. We describe the soliton-to-soliton stability limits of the free-carrier plasma-induced spectral broadening via single shot ADFT characterization in real time. In the presence of the free carrier dynamics, the soliton spectrum histograms demonstrate near-Gaussian limited intensity distributions with markedly high stability. Third, we investigate that free-carrier-perturbed modulation instability occurs in *both* normal and anomalous dispersion, has a broader gain spectrum and is thresholdless compared to the conventional Kerr modulation instability. Our studies are matched with the resonance condition of the Cherenkov radiation and modified modulation instability gain spectrum, including numerics and analytical solutions.

**Materials and Methods**

Figure 1a shows the hexagonal-lattice photonic crystal fabricated with optimized photolithography in a foundry. The device is 250 nm thick and the bottom $SiO_2$ cladding is 3 μm. The photonic bandgap nanomembrane, from a beginning silicon-on-insulator wafer, has periodicity, *a*, of 431 nm and hole of 125 nm (0.290*a*) radius. A single row translational defect is introduced, creating a 1.5 mm W1 slow-light photonic crystal waveguide. The slow-light silicon photonic crystal waveguide is then air-bridged by removing the bottom $SiO_2$ with buffered HF etch. The group-index $n_g$ and group-velocity-dispersion (GVD) of the sample are measured with an optically-clocked coherent swept wavelength interferometry (SWI) [25]. In the SWI, a fiber Mach-Zehnder interferometer with a 40 m path length imbalance is utilized to provide the timebase clock for an equidistant optical frequency sampling. In addition, a transmission spectrum of the hydrogen cyanide gas cell is acquired simultaneously as an absolute wavelength calibration to get around synchronization errors and ensure proper alignment of successive datasets. The resulting GVD and third-order dispersion (TOD) are then extracted from this delay-spectrogram and shown in Figure 1a.

The pump pulse we use to interrogate the slow-light soliton dynamics is a 39 MHz repetition rate laser with 2.3 ps full-width half-maximum pulse width ($\approx$ transform limited), and the center wavelength of the fiber laser is adjustable between 1545 and 1565 nm. A high fidelity second-harmonic-generation (SHG) based XFROG system (Figure 1b) is constructed to simultaneously capture the output pulse intensity, phase and spectrum [33-34]. Due to the presence of a pre-defined and strong reference pulse, the temporal direction degeneracy is avoided in the XFROG measurement and it can be used to characterize weak pulses with sub-fJ sensitivity. In our setup, half of the pump pulse is tapped out before the photonic crystal waveguide and utilized as the reference pulse in the XFROG system. The temporal structure of the pump pulse is independently characterized *a priori* using a standard SHG FROG measurement. The split and residual pump pulse is coupled into the photonic crystal structure. The pulse output, carrying information of the slow-light soliton dynamics, is treated as the test pulse of the XFROG system. As shown in Figure 1b, our measurements avoid any pulse amplification, pulse distortion and optoelectronic conversion to ensure high fidelity soliton characterization. Figure 1c shows an example XFROG spectrogram of the output pulse and the retrieved temporal pulse shape. To ensure proper convergence, the phase retrieval routine is concluded only when the FROG error drops below $5\times10^{-3}$ for all cases reported here [34].

Furthermore, we set up a single shot spectrum measurement system based on ADFT to study the shot-to-shot fluctuations of sub-picosecond slow-light soliton spectra at the repetition rate of 39 MHz [35-36]. In the ADFT setup, a long spool of dispersion-compensating fiber (DCF) encompassing normal dispersion of 1.33 ns/nm is used to provide the necessary time stretching as shown in Figure 1d, followed by a pre-amplifier along with a fast near-infrared photodiode-oscilloscope subsystem. The ADFT has an equivalent spectral resolution of 190 pm and its fidelity is confirmed by comparing the averaged measurements with the optical spectrum analyzer spectra. Before recording time series of up to 4,000 spectra, an averaged time-stretched spectrum (100 times) is captured and used as the reference spectrum for the cross-correlation analysis. The cross-correlation function is calculated using the definition:

$$C_{corr}(\tau) = \frac{\int r(t)s(t-\tau)dt}{\sqrt{\int |r(t)|^2 dt \int |s(t)|^2 dt}} \tag{1}$$

where $r(t)$ denotes the reference spectrum and $s(t)$ denotes the single-shot spectrum.

To help gain physical insights of the slow-light soliton dynamics, we adopt the symmetric Fourier split step method to compute the non-integrable nonlinear Schrödinger equation (NLSE) with an auxiliary equation describing the free carrier generation [26, 31]:

$$\frac{\partial E}{\partial z} = -\alpha E + i\sum_{k \geq 2} \frac{\beta_k}{k!}(i\frac{\partial}{\partial t})^k + i(s^2 \gamma_{eff}|E|^2 - sk_0 k_c N_c)E \tag{2}$$

$$\alpha = \frac{\alpha_{linear}}{2} + \frac{s^2 \beta_{TPA}}{2A_0}|E|^2 + sN_c \frac{\sigma}{2} \tag{3}$$

$$N_c = \int_{-\infty}^{T} \left(\frac{s^2 \beta_{TPA}}{2h\nu_0 A_0^2}|E|^4 - \frac{N_c}{\tau_f}\right)dt \tag{4}$$

Here $s = n_{eff}/n_o$ is the slow-light factor, $\beta_k$ the $k$-th order dispersion, $\alpha_{linear}$ the photonic crystal waveguide linear loss, $\gamma_{eff} = 2\pi n_2/\lambda A_o$ is the Kerr nonlinear coefficient, $\beta_{TPA}$ is the two-photon absorption (TPA) coefficient, $N_c$ is the TPA generated free-carrier density, $\tau_f$ is the free-carrier lifetime, $\sigma$ is the free carrier absorption coefficient (FCA), and $k_c$ is the dispersion coefficient of the free carriers (FCD). Parameters used are illustrated in Table 1 and are from either our measurements (linear loss and dispersion of the PhC waveguide) or published literatures [26, 31, 37]. Precise input pulse shape is directly retrieved from FROG measurement, providing exact initial conditions for the NLSE modeling.

**Results and Discussion**

Figure 1a plots the measured wavelength-dependent transmission and group index (upper panel), as well as the GVD and TOD of the PhC waveguide (lower panel). Within the PhC bandgap between 1540 nm and 1560 nm, the group index increases from 7.5 to 14.5 while the anomalous GVD can still support picosecond solitons. In this paper we highlight the most heuristic soliton evolution results pumping at 1555 nm ($n_g = 10.0$, $\beta_2 = -1.6$ ps$^2$/mm, $\beta_3 = 0.22$ ps$^3$/mm) and 1560 nm ($n_g = 14.5$, $\beta_2 = -4.5$ ps$^2$/mm, $\beta_3 = 1.0$ ps$^3$/mm).

Figure 2 shows the retrieved pulse profiles and spectra as a function of incident pulse energies. At 1555 nm where the slow-light factor is relatively small, the output pulse exhibits flat phase profile and temporal compression to 2.1 ps. On the other hand, with increasing slow-light factor at 1560 nm, soliton compression is absent but the output pulse exhibits strikingly large temporal acceleration and self-frequency blue-shift. The temporal acceleration and the self-frequency blue-shift can be tuned continuously by changing the input pulse energy (Figure 3a), and the maximum measured temporal acceleration is as large as 6.4 ps, almost three times the input pulse width.

These measurements deterministically captured the pulse dynamics with combined temporal and spectral information and thus offer ideal platforms to understand the increased interactions in the large group index regime. Consequently we model the pulse dynamics using the full-scale NLSE that includes slow-light scaling, exact device dispersion, TPA and dynamics of the free-carriers. Slow-light enhancements in the Kerr nonlinearity and TPA, both of which are rooted in third-order nonlinearity, scale with $s^2$ ($s$ denotes the slow-light factor), due to the prolonged light-matter interaction time and increased light intensity due to spatial compression [27, 37]. The slow-light enhancements of FCD and FCA are, however, more subtle. First, the generated electron and holes are from TPA and the carrier density scales quadratically with the laser intensity, which inherently scales with $s^2$ due to intensity enhancement [37]. Second, the accumulation time of free carriers ($T$ in Eq. 4) and the polarizability of free carriers ($s$ in the last term of Eq. 3) are both affected from the prolonged interaction time, leading to another $s^2$ enhancement. This can thus give rise to an overall $s^4$ scaling of FCA and FCD. Such mechanisms can be viewed under the interpretation that the time-accumulated free-carrier generation and the cascaded FCA and FCD provides more intense enhancement due to slow-light, than the Kerr nonlinearity which is based on the instantaneous electric polarizability [38].

As shown in Figure 2a to 2d, when $s^4$ scaling of FCD and FCA are included in the model, remarkable agreements are achieved between theory and experiment, in the soliton intensity shape, phase, spectrum and acceleration. Note that the values of $s$ for the wavelengths examined here ($s = 2.9$ for 1555 nm; $s = 4.2$ for 1560 nm) already reflect a ~70 to ~310× scaling of Drude free-carrier plasma effects, which are sufficient to manifest the underlying physics. To understand the slow-light scaling, we show in Figure 2e and 2f the root-mean-square-deviations between experimentally measured and NLSE calculated pulse intensity profile and spectra under different scaling rules. Unambiguously, with $s^4$ enhancement for free carrier effects, $s^2$ for Kerr and TPA respectively, the smallest deviation of temporal waveform and spectra is obtained for a diverse array of soliton energies. This is especially at the highest pulse energies that encounter largest nonlinear contributions, supporting the quartic $s^4$ enhancement scaling of the Drude plasma nonlinearities in slow-light photonic crystals.

Furthermore, numerical modeling clearly shows that both the temporal acceleration and self-frequency blue-shift originate from the slow-light enhanced Drude free-carrier effect. Compared with instantaneous Kerr effect, the nonlinear phase shift from FCD results from accumulation of free-carrier generated from TPA and thus has a quasi-linear profile as illustrated in Figure 3b. Consequently, the quasi-linear phase shift leads to a center frequency shift and the temporal acceleration follows due to the anomalous dispersion. In addition, the cascaded nonlinearity nature of the TPA-FCD process lends itself to more slow-light enhancement and thus the effect becomes more prominent at 1560 nm when the slow-light factor is higher. Moreover, it is seen in Figure 3c that increasing the 1555 nm pulse energy to 25 pJ improves the pulse acceleration up to 6 ps (more than 2.5× the pulse width) while maintaining favorable pulse shape with modest compression. Further increases in the pulse energy split the soliton pulse.

To further understand the quartic Drude slow-light scaling, we next examine the soliton perturbative analysis under TPA-FCD. With the detailed derivation in Supplementary Information Section I [39], the results can be summarized into the following two closed-form equations on the self-frequency blue-shift and temporal acceleration:

$$\Delta\omega = \frac{4}{15}\frac{k_0 k_c \beta_{TPA}}{h\nu_0 A_0^2}z \qquad (5)$$

$$\Delta t = -\frac{2}{15}\frac{k_0 k_c \beta_{TPA}|\beta_2|}{h\nu_0 A_0^2}z^2 \qquad (6)$$

When the system parameters (Table 1) are entered, the estimated temporal acceleration from Eq. 6 is 7.6 ps, matching well with the experimentally measured acceleration of 6.4 ps for 1560 nm pulse at 6.8 pJ. Of note, the effects of FCD shown in Eq. 5 are opposite to the self-frequency red-shift induced by Raman in nonlinear optical fiber [23] platforms. While Raman is more prominent when the system is interrogated by ultrashort pulses, the TPA-FCD effect is not changed with input pulse duration if the fundamental soliton condition is satisfied. Hence, relatively long pulses like the 2.3 ps pulses used in this work are preferred to observe the TPA-FCD effect in the absence of Raman.

Drude FCD also results in perturbation to the spectral broadening and optical Cherenkov radiation as depicted in Figure 4. Figure 4a shows experimentally measured spectral dynamics of 780 fs (FWHM) pulse. To facilitate nonlinear spectral broadening of the femtosecond pulses, we utilize a photonic crystal waveguide with specifically designed low GVD regime near 1545 nm via symmetric lattice shift [40-41] (GVD curve of this waveguide differentiates slightly from the typical W1 photonic crystal waveguide shown in Figure 1a, see details in Supplementary Information Section III [39]). Here the center wavelength is chosen at 1560 nm with $n_g \sim 14.0$ to harvest more slow light enhancement of FCD. As the input pulse energy is increased from 3.5 pJ to 7.0 pJ, the 10-dB bandwidths are increased from 12.5 nm to 15.4 nm, 22.2 nm and 25.1 nm respectively. The measurements (blue solid lines) are well-matched to the NLSE numerical modeling (red solid lines). To identify the origin of the observed spectral broadening, we artificially turn off the FCD in the NSLE numerical modeling and the magnitude of spectral broadening is greatly reduced (green dashed lines). While the Kerr nonlinearity is quenched by the nonlinear absorption in the silicon waveguide, Drude TPA-FCD takes over and dominates the spectral broadening in our slow-light waveguide parameters.

To investigate the Drude free-carrier plasma-perturbed dynamics of Cherenkov radiation, Figure 4b next shows the NLSE simulated spectral broadening kinetics in our waveguide with a 7 pJ pulse energy and ≈ 780 fs pulsewidth. The FCD contribution is numerically tuned to evaluate its cumulative impact on the output spectra. Particularly, without FCD, an isolated Cherenkov radiation sideband is obtained around 1534.0 nm (≈ 50 dB smaller than soliton spectrum), based on the zero-crossing of dispersion near 1545 nm. When the FCD effect is numerically tuned larger, a clear blue shift in the spectra is observed. Correspondingly, the dispersive wave manifests a red shift to counterbalance, cumulating in an overlapped soliton and dispersive wave spectra. Mechanisms for such dynamics are multifold: first, a blue-shift of soliton spectrum via FCD corresponds to a red-shift of phase-matched radiation spectrum [18]. Second, the blue-shifted soliton spectral overlaps better with the red-shifted radiation components, leading to rapid growth of radiation intensity [16]. Third, FCD modifies the phase-matching between soliton and dispersive wave, enabling contributions to both the wavelength and intensity of dispersive waves. Particularly, the phase-matching condition of the Cherenkov radiation [6,18] is now modified by the FCD to:

$$\sum_{n\geq 2}\frac{\beta_n(\omega_S)(\omega_D-\omega_S)^n}{n!}=\phi_{SPM}+\phi_{FCD} \qquad (7)$$

where $\beta_n$ is the $n$th order dispersion, $\omega_S$ and $\omega_D$ are the angular frequency of soliton and dispersive wave emission, and $\phi_{SPM}$ and $\phi_{FCD}$ are the phase shift due to SPM and FCD. Figure 4c shows the calculated phase-matching wavelengths with different values of the total nonlinear phase shift ($\phi_{SPM}+\phi_{FCD}$). First, with zero FCD and soliton wavelength at 1560.0 nm (green spot 1 in Figure 4c), the analytically estimated phase-matching wavelength is about 1534.0 nm (yellow spot 1 in Figure 4c), agreeing well with the NLSE simulations (green and yellow spot 1 in Figure 4b, see Supplementary Information Section II [39]). With increasing FCD, the Drude soliton spectrum undergoes a blue-shift (i.e., $\omega_S$ become bigger) while the phase shift from FCD becomes bigger (i.e., $\phi_{SPM}+\phi_{FCD}$ equals approximately to $2s^2\gamma_{\text{eff}}P_o$ and $3s^2\gamma_{\text{eff}}P_o$ for spot 2 and 3 respectively), and the estimated resonance conditions are marked in Figure 4c (spot 2 and 3), also consistent with the numerical simulations in Figure 4b (spot 2 and 3 correspondingly) and our qualitative analysis above. Subsequently, with FCD fully applied, soliton spectrum and Cherenkov radiation emerge together. Therefore, in spite of that Cherenkov radiation is not distinctly separated from soliton spectra, our analysis points out the underlying contribution and dynamics of the Cherenkov radiation perturbed by the Drude FCD. Furthermore, the FCD-induced Cherenkov radiation beneath soliton spectrum is verified with the modelled time-domain waveform illustrated in Figure 4d. Particularly, with FCD applied, a larger but more slowly oscillating dispersive wave tail is obtained due to the overlapping of Cherenkov and soliton spectrum, in contrast to the smaller and faster dispersive wave tail incurred by the far separated Cherenkov and soliton spectrum without the Drude FCD.

With the spectral broadening induced by slow-light enhanced FCD, ADFT is next implemented to investigate the soliton-to-soliton stability of the broadened Drude soliton spectral components. Figure 5a shows the averaged optical spectrum of the spectrally broadened pulses, measured with a grating based optical spectrum analyzer (OSA). Single-shot ADFT measurements are conducted not only on the pump at 1560 nm but also around 1540 nm, especially the spectral corners wherein unstable features should be expected if any [42]. Figure 5b shows the recorded single shot spectrum waveforms. ADFT not only fully retrieves the detailed spectral characteristics, but also unveils the real-time spectral dynamics that cannot be acquired in conventional optical spectrum analyzers or RF spectrum analyzers. Furthermore, Figure 5c summarizes the cross-correlation of the time-averaged (reference) spectrum from the OSA and the single-shot spectra, where nearly Gaussian spread around +1 containing minuscule instability are observed. The computed cross-correlation average values are $M_{fcd}$ = 0.9989 and $M_{pump}$ = 0.9978 for the FCD-sideband and near the pump respectively. Both values of variation ($V_{fcd}$ = 3.4×10$^{-8}$ for FCD sideband and $V_{pump}$ = 2.0×10$^{-7}$ near the pump) are small and within the instrumentation measurement noise floor. This demonstrates the stable soliton-to-soliton behavior of Drude TPA-FCD induced spectral broadening, capturing the high stability of the Drude free-carrier nonlinear absorption ($\sigma$) and dispersion $k_c$ parameters.

To further examine soliton dynamics under Drude free-carrier plasma, we next examine theoretically and numerically the effects of FCD onto the modulation instability (MI) gain spectrum. We assume:

$$E = [\sqrt{P} + a(z,t)]e^{i\left[\gamma_{eff}P - \frac{k_0 k_c \beta_{TPA}\tau_f}{2h\nu_0 A_0^2}P^2\right]z} \qquad (8)$$

$$N_c = \frac{\beta_{TPA}\tau_f}{2h\nu_0 A_0^2}P^2 + b(z,t) \tag{9}$$

*a* and *b* are the perturbations. Here the implicit assumption is that the absorption is not yet too strong such that we can assume the steady-state solution has constant amplitude along the propagation. This is a necessary approximation because no analytic steady-state solution exists if we consider all the dynamical absorptions. Using linear stability analysis, we obtain the Drude-perturbed MI gain spectrum as (detailed in Supplementary Information Section IV [39]):

$$G_{MI} = \text{Im}\left(|\beta_2 \Omega| \sqrt{\Omega^2 + \frac{4\gamma_{eff}P}{\beta_2} - \frac{4k_0 k_c \beta_{TPA} P^2}{h\nu_0 A_0^2 \beta_2 \left(\frac{1}{\tau_f} - i\Omega\right)}}\right) \tag{10}$$

It is seen from Eq. 10 that, different from conventional Kerr MI utilized in optical parametric oscillators [23, 43-44], MI gain is always guaranteed by Drude FCD regardless of MI frequency and present in both normal and anomalous group velocity dispersion. Figure 6a shows the numerically calculated MI spectra with a 3W continuous-wave pump. Agreeing with our analysis, the MI induced by Drude FCD features a thresholdless amplification and a broader gain bandwidth. Furthermore, as shown in Figure 6b, in the normal dispersion region, FCD-perturbed MI is possible, fundamentally different from canonical Kerr MI which only exists in the region of anomalous group velocity dispersion. In both regions, the spectra drop off as the increase of frequency detuning due to the 1/Ω term in Eq. 10 [45]. Moreover, it is seen from Eq. 10 that MI perturbed by FCD is affected by free-carrier lifetime (the last term under the square root of Eq. 10), which is confirmed by numerical simulations shown in Figure 6c. In particular, with decreased $\tau_f$, the numerically calculated MI spectra become narrower, consistent with the trend predicted by Eq. 10. With zero free-carrier lifetime, the contribution from FCD totally vanishes and the numerical MI spectrum turns into the pure Kerr case. Finally, Figure 6d shows the influence of TPA on MI gain spectrum. Particularly, for MI induced by Kerr effect, TPA loss severely suppresses the MI sidebands because Kerr MI originates from parametric amplification that is directly proportional to light intensity. However, for FCD-induced MI, if TPA is switched off, the free carrier densities goes to zero, collapsing the MI gain spectrum back to its original spectrum. The uniqueness of FCD-induced MI dynamics holds potential for realizing on-chip broadband laser sources.

## Conclusions

Here we examined ultrafast solitary wave kinetics in slow-light silicon-based photonic bandgap waveguides based on ADFT real-time spectrum analysis and phase-resolved XFROG. With slow-light enhanced Kerr nonlinearity and Drude free-carrier effects, we demonstrated a quartic slow-light scaling of the Drude free-carrier kinetics. Our measurements and analytical solutions captured with high fidelity the self-induced soliton acceleration, the self-frequency soliton blue-shift, and the asymmetric spectral broadening under 780 fs pulse excitations. We further extended the Drude plasma perturbation model to broadband modulation instability, illustrating a thresholdless MI gain spanning across the regions of normal and anomalous group velocity dispersion. Drude dynamics also modified the resonance condition for Cherenkov radiation. To uncover the stability limits of Drude dynamics, we subsequently examined the real-time and single shot soliton-to-soliton dynamics through ADFT. With the Drude plasma, the soliton histograms illustrate near-Gaussian limited distributions with pump and spectral sideband cross-correlations at the instrumentation noise limits. These observations advance fundamental insights on the combinational Drude plasma and Kerr dynamics of picojoule solitons in chip-scale silicon dispersive media.

## Acknowledgements

The authors acknowledge discussions and contributions with Pierre Colman, Bahram Jalali, Stefano Trillo, Jiangjun Zheng, Tingyi Gu, Jinghui Yang, Hao Zhou, Richard M. Osgood Jr., Chad A. Husko, Matthew D. Marko, Pin-Chun Hsieh, Jiali Liao, and Jiankun Yang. Funding support is from Office of Naval Research with grant N00014-14-1-0041 and H.Z. acknowledges UESTC Young Faculty Award ZYGX2015KYQD051 and the 111 project (B14039). X.L. acknowledges funding from NSFC Grant 61070040. S.W.H acknowledges funding from AFOSR Young Investigator Award with grant FA9550-15-1-0081.

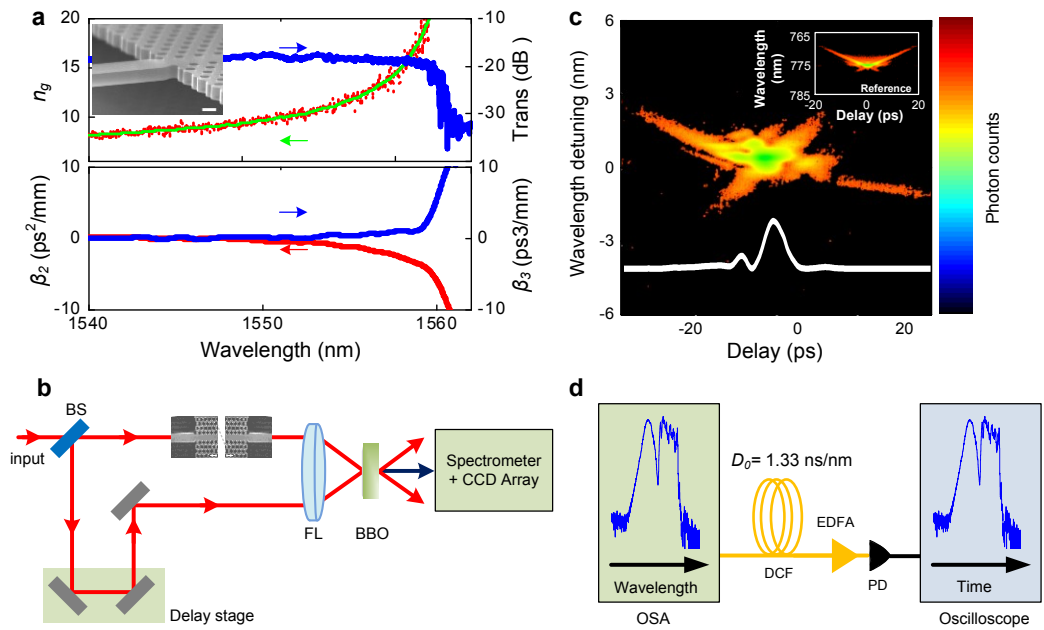

**Figure 1 | XFROG and real-time single shot spectroscopy utilized to elucidate the Drude free-carrier soliton dynamics. a,** Group indices (data points with fitted green solid line), group-velocity-dispersion (red) and third-order-dispersion (blue) of the slow-light photonic crystal (PhC), measured via optically-clocked swept wavelength interferometry. The waveguide transmission is shown in the upper panel blue curve. Inset: SEM of the PhC. Scale-bar length is 400 nm. **b,** XFROG approach using second harmonic. BS: beam splitter, FL: focusing lens, BBO: barium borate nonlinear crystal. **c,** Illustrative XFROG spectrogram of the 1555 nm soliton at the output of the slow-light PhC waveguide. The retrieved intensity waveform is shown in white. Inset: reference FROG spectrogram for the soliton measurements. **d,** ADFT setup for real-time soliton characterization. DCF: dispersion compensation fiber, PD: photodiode, OSA: optical spectral analyzer.

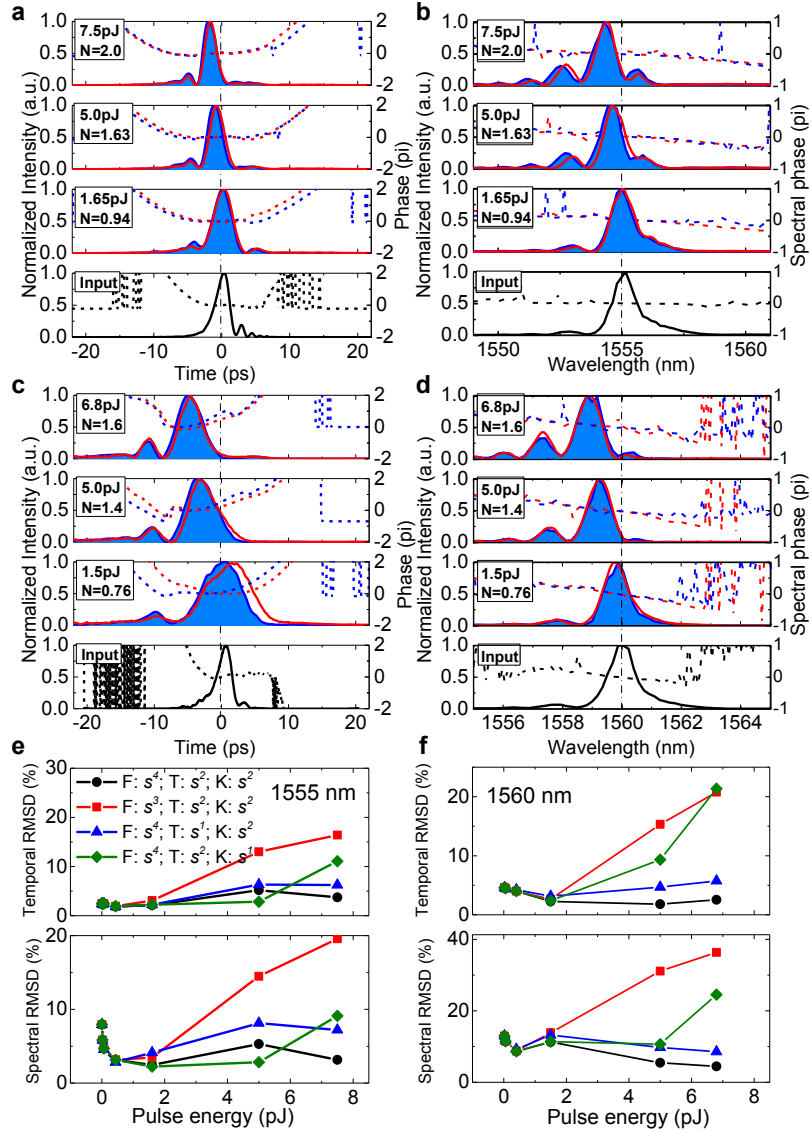

**Figure 2 | Quartic slow-light scaling of Drude free-carrier nonlinear dynamics on soliton propagation.** The solitary dynamics are mapped with different input pulse energies and soliton numbers $N$, for two example wavelengths of 1555 nm ($n_g = 10.0$) and 1560 nm ($n_g = 14.5$). **a** and **b**, For 1555 nm, the blue shades and blue dashed lines are the experimental cross-correlation retrieved intensity and phase respectively. The red dashed and red solid line plots are the captured phase and intensity from the numerical predictions. **c** and **d**, Likewise for 1560 nm. In both center-wavelength cases, soliton acceleration and self-frequency blue-shifts are observed at various magnitudes. **e** and **f**, Measurement comparison with the pulse evolution modeling via NLSE including the Drude free-carrier dynamics, elucidates the exponent in the slow-light scaling. Panel **e** is for 1555 nm soliton and **f** for 1560 nm soliton. The top panel is the temporal measurement-theory root-mean-square deviations (RMSD). The bottom panel is spectral measurement-theory RMSD. Letters in the legend are: F = FCD and FCA dynamics; T = TPA; K = Kerr nonlinearities.

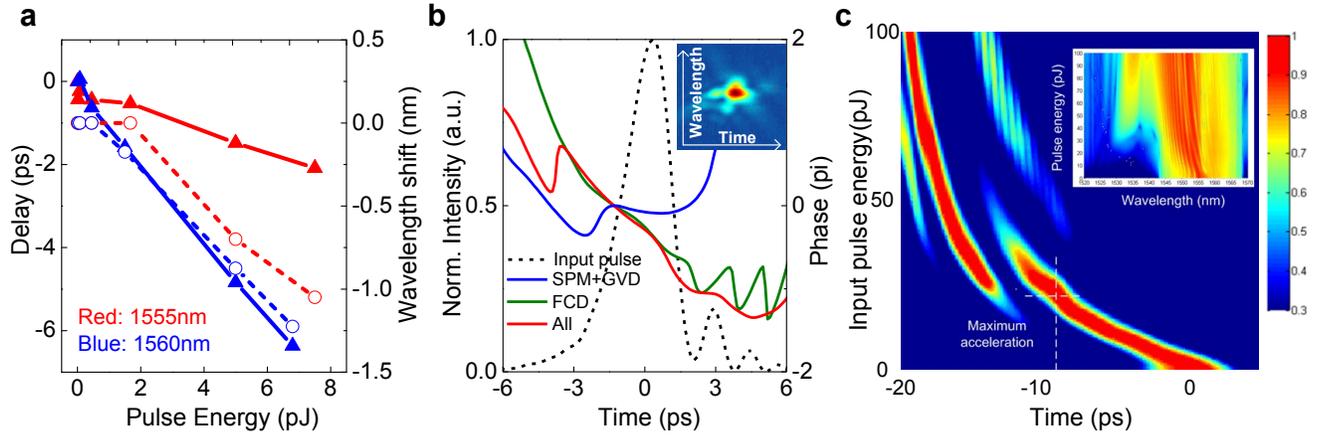

**Figure 3 | Influence of Drude free-carrier dynamics on the slow-light soliton dynamics. a,** Under the Drude free-carrier perturbation, the measured soliton accelerations and self-frequency blue-shift are shown in the solid line with triangles and dashed line with circles respectively. 1560 nm and 1555 nm slow-light examples are illustrated in blue and red. **b,** The modeled contributing phase shifts from Drude free-carrier dynamics, self-phase modulation and group velocity dispersion. Input is a 1555 nm pulse at 7.5 pJ ($N = 2.0$) with pulse shape illustrated in dashed background. Inset shows the corresponding measured XFROG spectrogram. **c,** For a 1.5 mm slow-light photonic crystal waveguide, the maximum acceleration with acceptable pulse shape is about 9.0 ps. Parameters used in the simulation are identical to Figure 2**a** and **b**. Insets show the corresponding spectral evolutions.

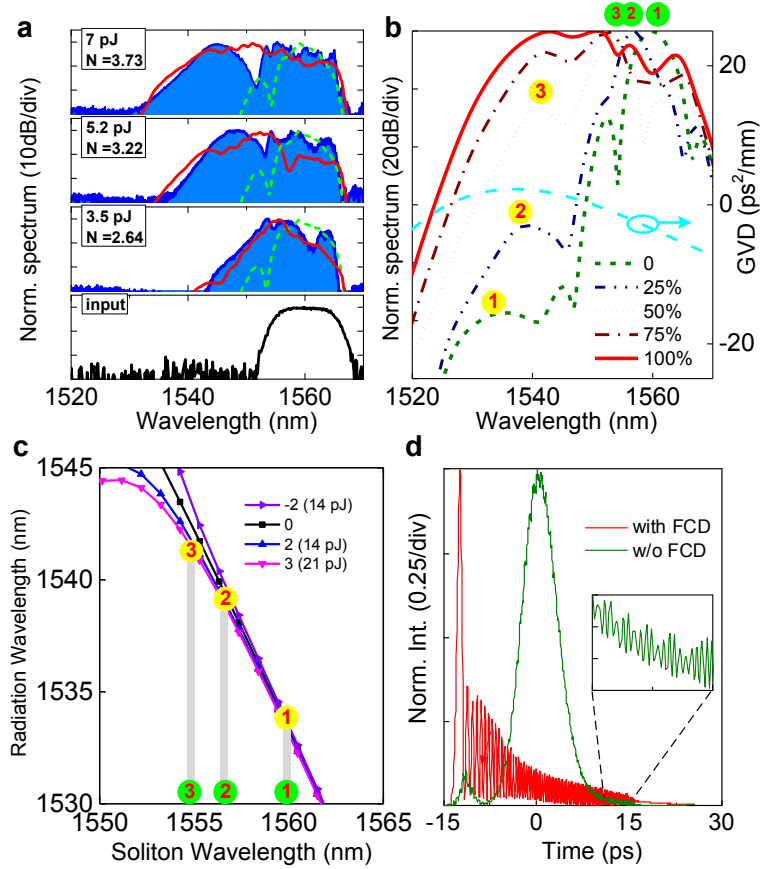

**Figure 4 | Drude FCD-based broadened spectra with optical Cherenkov radiation in the femtosecond soliton regime. a,** Experimental (blue) and the NLSE simulated (red) spectral broadening with different pump pulse energies and solitary wave number *N*. Excellent agreement is achieved when the quartic slow-light scaling is included in the analysis. For illustrative comparison, dashed green lines plot the NLSE simulated spectra when the FCD effect is numerically turned off. **b,** NLSE simulated spectra and dynamics from Cherenkov radiation with increasing Drude free-carrier dispersion (FCD) coefficient contributions. The estimated Cherenkov radiation peaks at yellow spot 1 to 3 are: 1534.0 nm, 1538.5 nm and 1541.0 nm; and the corresponding soliton center wavelengths at green spot 1 to 3 are: 1560.0 nm, 1556.7 nm and 1554.8 nm. Dispersion curve for the simulation is shown to the right vertical axis, with $\beta_2 = -3.5$ ps$^2$/mm, $\beta_3 = 0.5$ ps$^3$/mm, $\beta_4 = -0.005$ ps$^4$/mm, and $\beta_5 = -0.0025$ ps$^5$/mm, centered at 1560 nm. The zero crossing of dispersion comes from symmetrical lattice shift to build low GVD region within the passband of the photonic crystal waveguide. **c,** Analytically estimated phase-matching condition for Cherenkov radiation under different effective nonlinear phase shift (corresponding pulse energy in the brackets). The nonlinear phase shifts are normalized to $s^2 \gamma_{eff} P_o$. **d,** The temporal waveform corresponding to the spectra shown in panel **b**, with FCD present and FCD absent. Inset: the close-up signal of the temporal dispersive wave tails without FCD, illustrating the faster and weaker temporal oscillations that arise from the heterodyne beating of the far separated dispersive wave from the soliton.

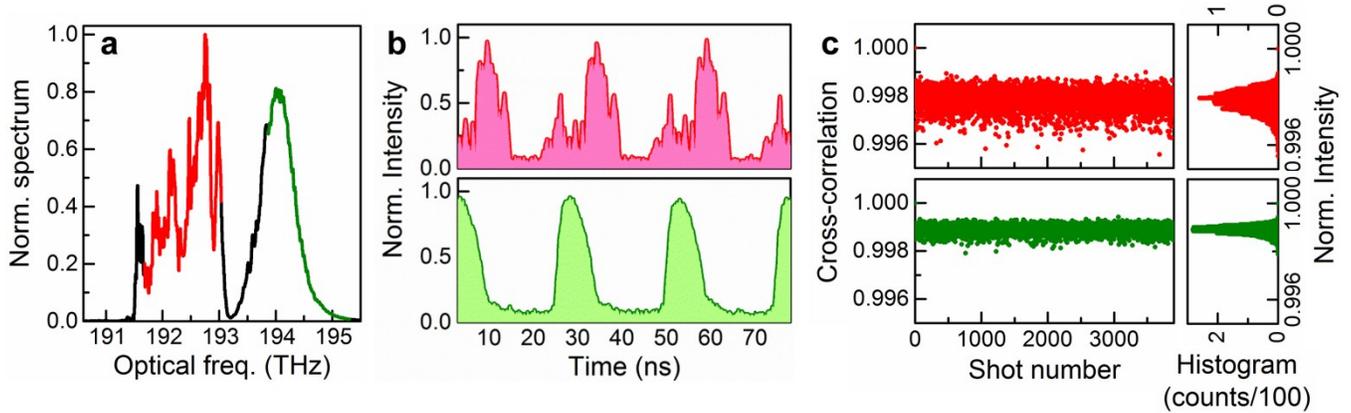

**Figure 5 | ADFT based single shot and real-time spectral measurements. a,** Averaged spectrum of the Drude FCD-induced spectral broadening and optical Cherenkov radiation, measured with an OSA. The soliton pulse energy is 7.0 pJ and the center wavelength is 1560 nm. **b,** Example single shot spectra corresponding to the red (near the pump wavelength) and green (the edge of the broadened spectrum) highlighted spectral components respectively. The traces are captured using a real-time oscilloscope, with the temporal horizontal axis swapped to optical spectra through ADFT. **c,** The single shot spectra and time-averaged spectrum cross-correlation. Histograms of the cross-correlations demonstrate the nearly Gaussian spread around +1 with minuscule variations. Average cross-correlation for Drude-broadened spectral sideband $M_{fcd}$ is 0.9989, with a variation $V_{fcd}$ of $3.4 \times 10^{-8}$. The cross-correlation mean value near the pump $M_{pump}$ is 0.9978, with a variation $V_{pump}$ of $2.0 \times 10^{-7}$. Both variation values are small and within the measurement noise floor, demonstrating stable soliton-to-soliton behavior.

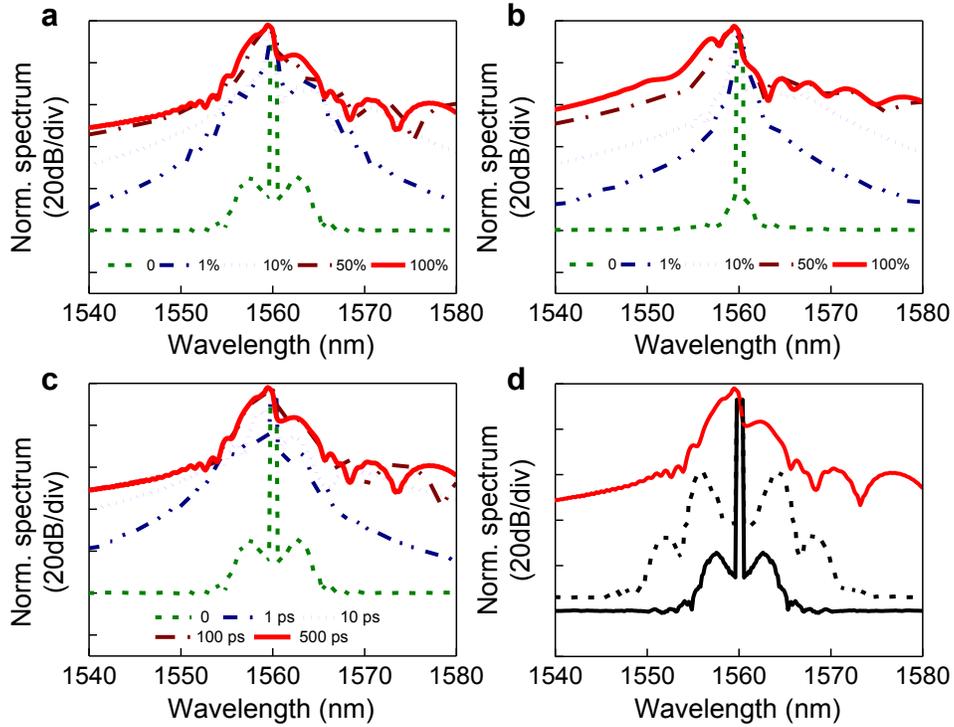

**Figure 6 | Theoretical study of concurrent Drude FCD-perturbed MI in silicon photonic crystal waveguides. a-b**, NLSE simulated FCD perturbed MI spectra of a 3W continuous wave pump for increasing FCD strength, panel **a** for anomalous dispersion regime ($\beta_2 = -2.2$ ps$^2$/mm), and panel **b** for normal dispersion regime ($\beta_2 = 2.2$ ps$^2$/mm). **c,** NLSE simulated FCD perturbed MI gain spectra for different free-carrier lifetime in anomalous dispersion regime ($\beta_2 = -2.2$ ps$^2$/mm). **d**, Comparison of the influence of TPA loss on MI gain spectrum with (red) and without (black) the Drude FCD for a 3W continuous wave pump in anomalous dispersion regime ($\beta_2 = -2.2$ ps$^2$/mm). Solid lines show the cases with TPA loss and the dotted line show the case without TPA loss. Note that the solid red line does not have a corresponding dotted red line considering that FCD must occur together with TPA.

**Table 1** | Parameters summary used in the NLSE simulations.

| Parameters | Values |
|---|---|
| linear loss $\alpha_{\text{linear}}$ | 2.5 dB/mm |
| Kerr nonlinear index $n_2$ | $5\times10^{-18}$ m$^2$/W |
| Effective nonlinear coefficient $\gamma_{\text{eff}}$ | $\dfrac{2\pi n_2}{A_0 \lambda}$ |
| two-photon absorption $\beta_{TPA}$ | $8.8\times10^{-12}$ m/W |
| free carrier absorption $\sigma$ | $1.45\times10^{-21}$ m$^2$ |
| free carrier dispersion $k_c$ | $1.35\times10^{-27}$ m$^3$ |
| modal area $A_0$ | 0.13 μm$^2$ |
| Free carrier lifetime $\tau_f$ | 500 ps |
| Soliton number $N$ | $\sqrt{\dfrac{s^2 P_0 \gamma_{eff} T_0^2}{|\beta_2|}}$ |